# Seamless Handover for IMS over Mobile-IPv6 Using Context Transfer

Reza Farahbakhsh, Naser Movahhedinia
Computer Engineering Department, Faculty of Engineering, University of Isfahan, Isfahan, Iran

**Abstract:** Mobility support for the next generation IPv6 networks has been one of the recent research issues due to the growing demand for wireless services over internet. In the other hand, 3GPP has introduced IP Multimedia Subsystem as the next generation IP based infrastructure for wireless and wired multimedia services. In this paper we present two context transfer mechanisms based on predictive and reactive schemes, to support seamless handover in IMS over Mobile IPv6. Those schemes reduce handover latency by transferring appropriate session information between the old and the new access networks. Moreover, we present two methods for QoS parameters negotiations to preserve service quality along the mobile user movement path. The performances of the proposed mechanisms are evaluated by simulations.

## I. INTRODUCTION

Third Generation Partnership Project (3GPP) has presented IP Multimedia Subsystem (IMS), to support multimedia services such as voice, video, real-time and interactive applications over IP networks. One key feature that makes IMS a promising technology is its independency to access mechanisms such that IMS services can be provided over any IP connectivity networks (e.g., UMTS, WiFi, x-DSL, and WiMAX). The IMS connection initiation procedure consists of network elements used in Session Initiation Protocol (SIP) based session control[1]. The SIP is employed at the application-layer in order to simplify integrating IMS with the Internet protocols. Furthermore, IMS represents reference service delivery platform architecture for the IP multimedia services within an emerging mobile all-IP network environment.

Recently IMS is considered to offer real-time IP multimedia applications over wireless mobile networks. As more bandwidth hungry and sensitive to error applications are emerging, the two important issues, resource provisioning and handover management, should be seriously addressed for such networks. The resource provisioning assures providing sufficient network resources to User Equipments (UEs) for Quality of Service (QoS) guarantees. The handover management enables a UE to keep the network connectivity during point of attachment alterations.

As a major trend, networks have been evolving from traditional circuit-switched architectures to an all-IP paradigm. Per se, IMS may be considered as the foundation for future wireless and wire line convergence. In order to benefit from the advantages of IPv6, 3GPP has selected it as the IP version supported by the IMS. Accordingly, Mobile IPv6 has become a global solution to support mobility between various access networks, and one of the fundamental characteristics of IMS is the support for user mobility. However, moving form one access network to another may call for the MN (Mobile Node) to change its IP address and some other session specifications. To provide seamless communications, the handover procedure has to apply specific mechanisms to preserve the session states.

In this paper, two context transfer solutions are proposed to transfer the session state between old and new P-CSCFs in order to reduce handover latency time. We analyze the handover latency using timing diagrams, and calculate signaling cost to evaluate the performance of our proposed schemes. Furthermore, using those session state transfer schemes, we present two methods for QoS parameters negotiation to facilitate resource reservation required for service quality provisioning along the UE movement path. The rest of this paper is organized as follows.

The next section presents a brief overview of the IMS architecture, MIPv6 Protocol, context transfer mechanisms, QoS provisioning in IMS. In Section III, we present our proposed handover schemes to decrease the handover latency. In Section IV, the proposed schemes are evaluated using timing diagrams and cost analysis. Section V provides numerical simulation results and Section VI presents our QoS proposed schemes, finally section VII concludes this paper.

## II. BACKGROUNDS AND RELATED WORKS

### 2.1 IP Multimedia Subsystem architecture

3GPP has considered a layered approach for IMS architectural design. As shown in Figure 1, IMS network comprises transport, session control, and application layers. The signaling plane of the IMS is comprised of several functions: CSCF (Call Ses-sion Control Function), HSS (Home Subscriber Server), gateway functions (Media, Breakout, and Signaling), service and application servers. The HSS is the main data storage for all the subscriber and service-related data. Networks with more than one HSS do require SLF (Subscription Locator Function) which is a simple database that maps user's addresses to HSSs. The databases in the IMS architecture are shown in Figure 1.

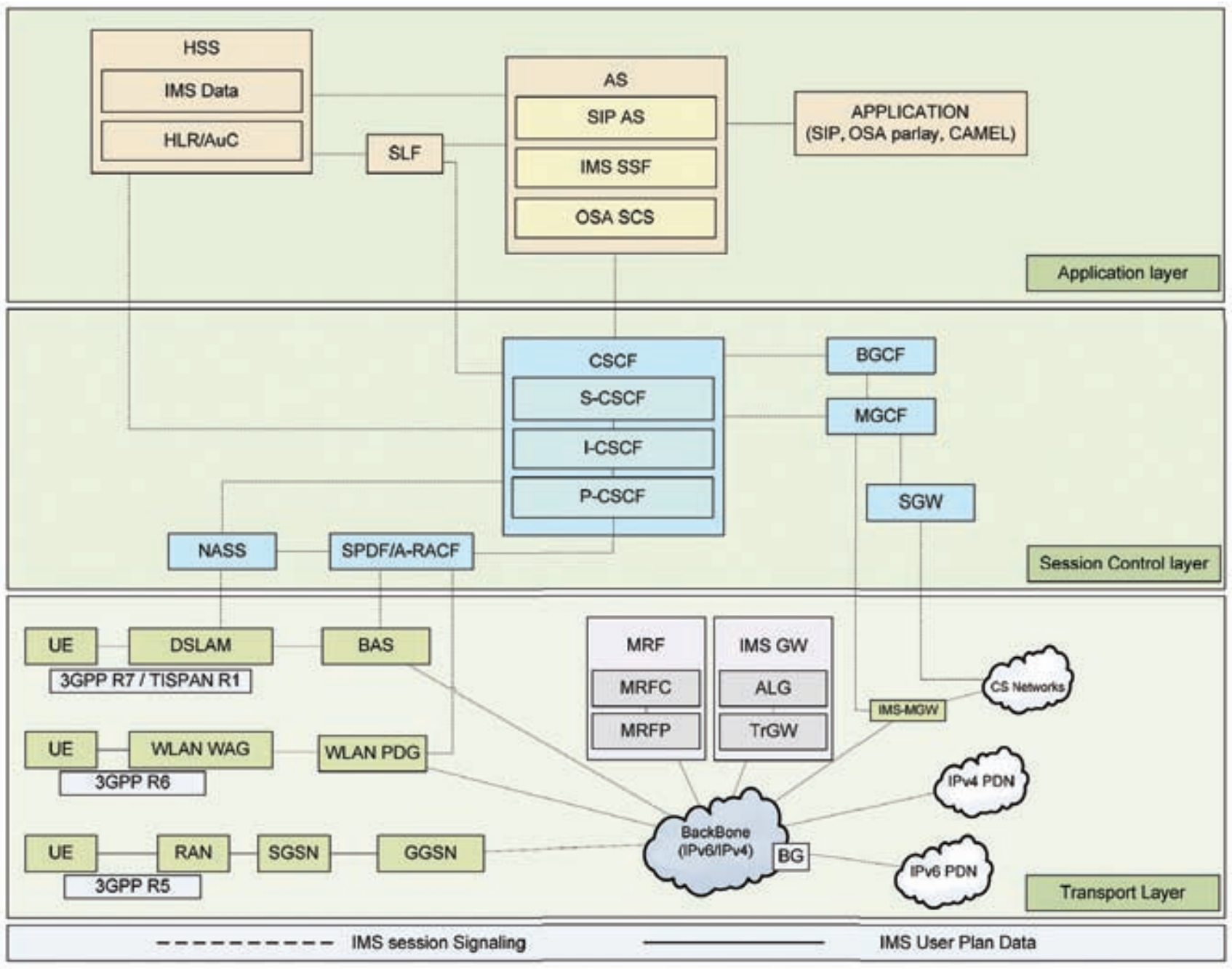

Fig.1 IP Multimedia Subsystem logical architecture

The heart of IMS is the CSCF that performs session control services. The CSCF is a SIP proxy, which handles the session establishment function by routing SIP messages that follow a pre-specified flow. Three types of CSCFs are defined; each type provides different functionality. Proxy CSCF (P-CSCF) is the entry point to IMS and is responsible for all communication with the home network. P-CSCF acts as the proxy between the user equipment (UE) and the Serving CSCF (S-CSCF) and sends all messages received from the UE to the S-CSCF. S-CSCF is the focal point of the IMS that performs the session control services for the IMS subscriber, maintaining session states and storing the service profiles. Interrogating CSCF (I-CSCF) is a contact point from any external network to the home network operator for an incoming session or for roaming vinteractions from a visited network.

The AS (Application Server) is a SIP entity that hosts and executes services. The AS nterfaces the S-CSCF using SIP. Also it has ability to send accounting information for charging functions.

The Media Resource Function (MRF) provides a source of media in the home network. It is further divided into a signaling plane node called the MRFC (MRF Controller) and a media plane node called the MRFP (MRF Processor). The BGCF (Breakout Gateway Control Functions) is essentially a SIP server that includes routing functionality based on telephone numbers. The BGCF is only used in sessions that are initiated by an IMS terminal and addressed to a user in a circuit-switched network, such as the PSTN or the PLMN. One or more PSTN gateways that contain SGW (Signaling gateway), MGCF (Media Gateway Controller Function) and MGW (Media Gateway), provide an interface toward a circuit-switched network, allowing IMS terminals to make and receive calls to and from the PSTN (or any other circuit-switched network).

To set up a session in IMS, the MN sends an Invite to the P-CSCF and addresses the correspondent node. The P-CSCF forwards the request to the S-CSCF based on the setup service route information. The P-CSCF verifies that the request was coming on a valid security association, whereas the S-CSCF trusts the requests coming from the P-CSCF since they are in the same trust domain.

### 2.2 Overview of mobile IPv6 protocol

Handoff management protocols can be broadly classified according to the layer of operation. Mobile IPv6 (MIPv6) is one of the most important protocols to accommodate the increasing demand of end-to-end mobility in IPv6 Internet[2]. The handover procedure in MIPv6 consists of movement detection, Duplicate Address Detection (DAD) for Care of Address (CoA) configuration, and Binding Update (BU).

There have been a lot of research and investigations to improve handover performance. FMIPv6[3] is a modification of MIPv6 that tries to reduce handover latency by utilizing Layer 2 triggers. Hierarchical MIPv6 (HMIPv6)[4] aims to reduce the signaling load due to user mobility. FMIPv6 reduces the handover delay by delivering packets in the new point of attachment at the earliest chance[1].

### 2.3 Context transfer procedure

Context Transfer Protocol (CXTP) is a network level protocol which transfers connection information and contexts between access routers[5].

It allows better node mobility support, and avoids re-initiation of signaling to and from an MN. Example features contained in the context are session information, AAA, QoS, security and header compression. The key goals are to reduce latency, minimize packet loss rate and avoid re-initiation of signaling to and from the MN. This mechanism is a handover optimization procedure to reduce the duration of interruption (handover delay) in user's ongoing application sessions[6].

Context transfer is used when the transmission path of a session changes and sessionrelated states are re-located from the network nodes on the old transmission path to the network nodes on the new transmission path. Context transfer is used, on one hand, to reduce the delay introduced by handover and on the other hand, to minimize the data loss during handover.

**2.4 Related works**

In [7] based on Context transfer procedure, a solution is introduced that reduces the handover delay and QoS provisioning in macro mobility by sharing the registration information and call states. In [8] a mechanism was presented for faster QoS establishment in SIP handover based on context transfer techniques. The proposed mechanisms decrease the volumes of signaling for the session re-establishment at the new access network and, therefore, lead to reduced handover delay.

To provide seamless handover between heterogeneous networks, [9] presents context transfer solution for fast handover in mobile IPv6 environment. Its results show that fast handover with context transfer at the network layer can support uninterrupted VoIP. [10] has proposed a framework for the end-to-end QoS context transfer in Mobile IPv6 which provides an end-to-end QoS context transfer for real-time applications and minimizes the handover service disruption, by avoiding end-to-end QoS signaling from scratch after handovers.

## III. HANDOVER PROCEDURE IMPROVEMENT

In this section, first we discuss the handover procedure which is used currently in IMSMIPv6 networks, and then propose two handover schemes using context transfer.

**3.1 Problem statement**

In recent years, there has been a rapid growth in the need to support moving hosts using mobile networks. Due to the mobility, the UE requires to alter the point of attachment to the access network. As the consequence the UE triggers a change of the MN's IP address and before continuing the session, it has to register to the IMS again. After the registration, the UE has to renegotiate parameters for the session before the session can be continued. Figure 2 shows the message flow of the Register and Invite for the IMS in presence of MIPv6[7 ~ 8, 11~14]. This will potentially introduce a long interruption of the ongoing session.

The standard MIPv6 handover timing diagram in an IMS network is shown in Figure 3.

**3.2 Proposed context transfer-based schemes**

We assume that a multimedia session is ongoing between two users. After some time, the MIPv6 node performs a handover from the access network where the session was generated to a new access network where the session should be continued. In this process, the MN changes its IP address which may imply a change of P-CSCF. According to the IMS specifications, the session at the old P-CSCF is terminated, and the MN has to trigger the standard SIP-based IMS procedures at the New P-CSCF (see Figure 4). New P-CSCF does not have any information about the MN and its session, so the MN has to register in it and re-invite the CN (see Figure 2).

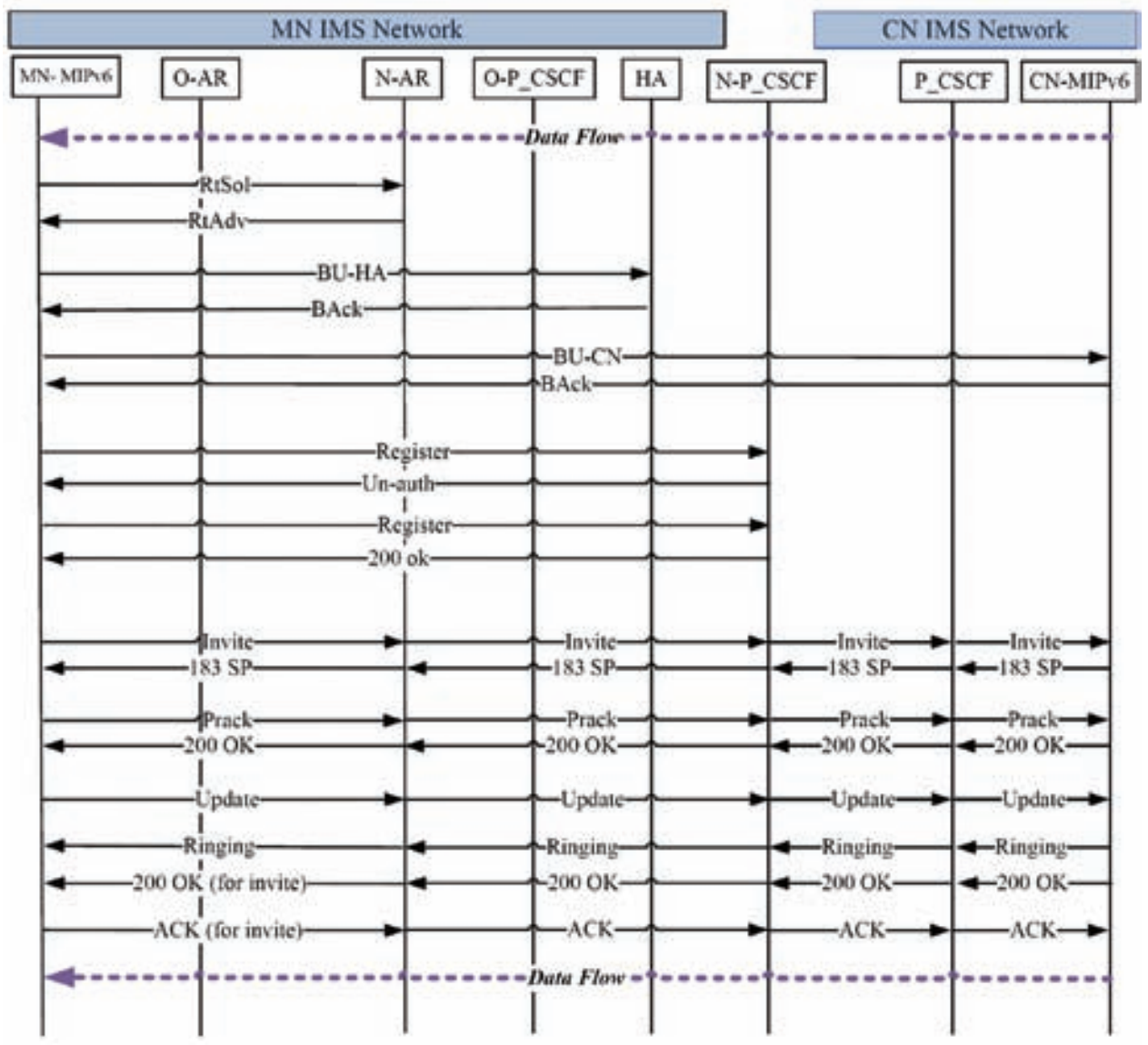

Fig. 2 Standard MIPv6 handover in IMS networks

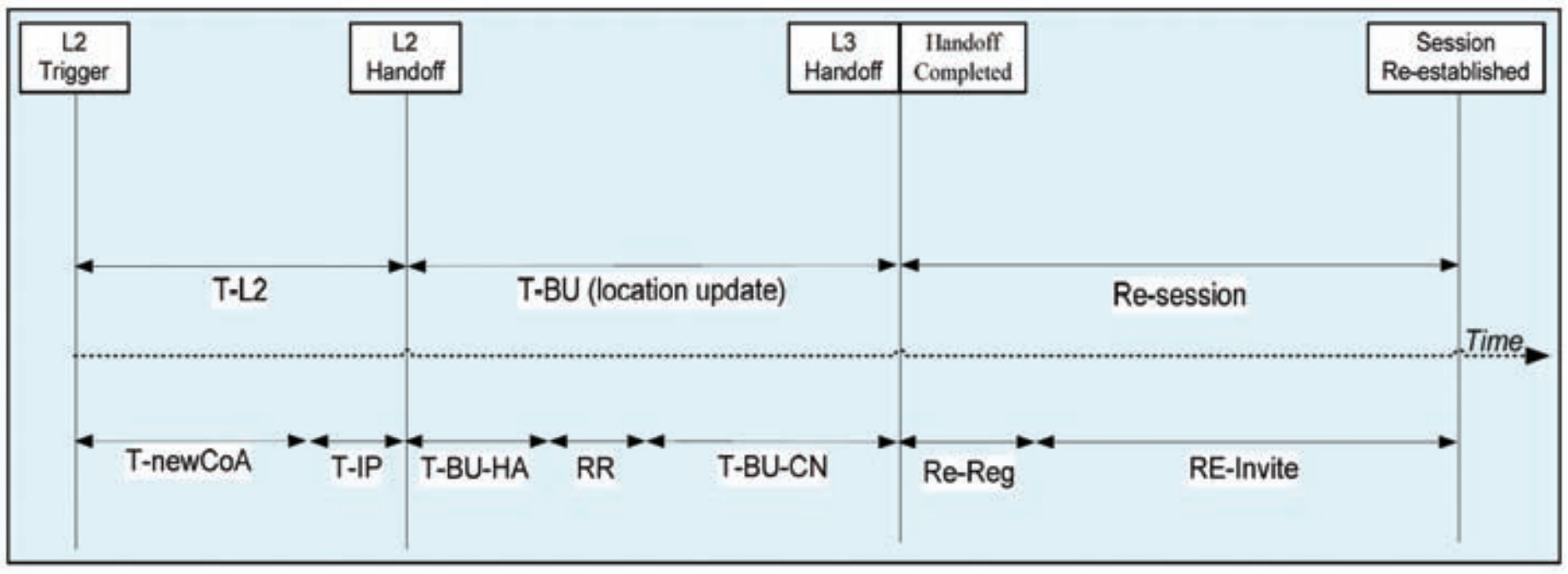

Fig. 3 Time diagram for Standard MIPv6 handover in IMS networks

The proposed solutions make it possible to have a handover between P-CSCFs, without losing session state information. Additionally, fewer messages are used for the re-register and re-invite of the MN and shorter handover delay is imposed for re-establishment of the session, comparing to the standard scheme. By transferring the session context (see Figure 4), new P-CSCF receives information about the MN and its session at the old

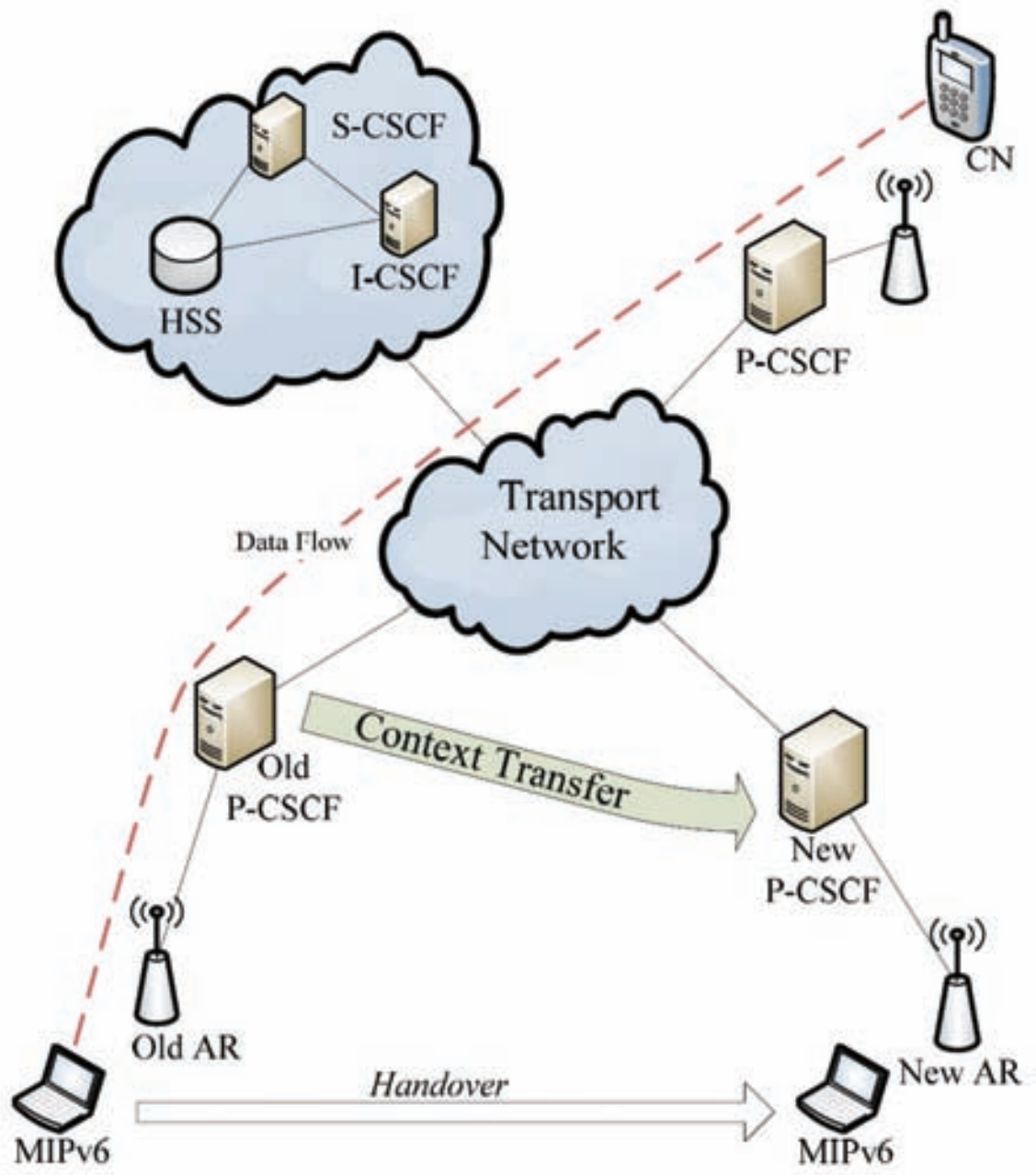

Fig. 4 Context transfer in proposed handover scenario

PCSCF, such as registration state, session states, Final Network Entry point, UE Address, Public and Private User IDs and Access Network Type.

### 3.3 Proposed scheme I

In this scheme, the MN knows in advance toward which router it will move and anticipates the transfer to the NAR. Note that this knowledge can be acquired by FMIPv6 and Neighbor Discovery (ND) [12]. The MN sends a Move-notify message to its P-CSCF (see Figure 5) containing the NAR address and the identifiers of the context to be transferred. Afterward, the old P-CSCF sends move-notify to the new P-CSCF and then transfers the context information to that. This information includes registration and session states, Final Network Entry point, UE Address, Public and Private User IDs and Access Network Type. After receiving Ack, old P-CSCF updates S-CSCF routes information by sending Route update and receives 200 ok. This proposed scheme utilizes movement anticipation, tunneling, and session context transfer to alleviate handover delay. It's worthy to mention that in this scheme the context transfer procedure is performed prior or simultaneous to MIPv6 handover and doesn't cause excessive latency.

The Proposed Predictive FMIPv6 handover Timing Diagram in IMS network is shown in Figure 6. As mention before, in this scheme the context transfer procedure is performed prior or simultaneous to MIPv6 handover and doesn't cause excessive latency.

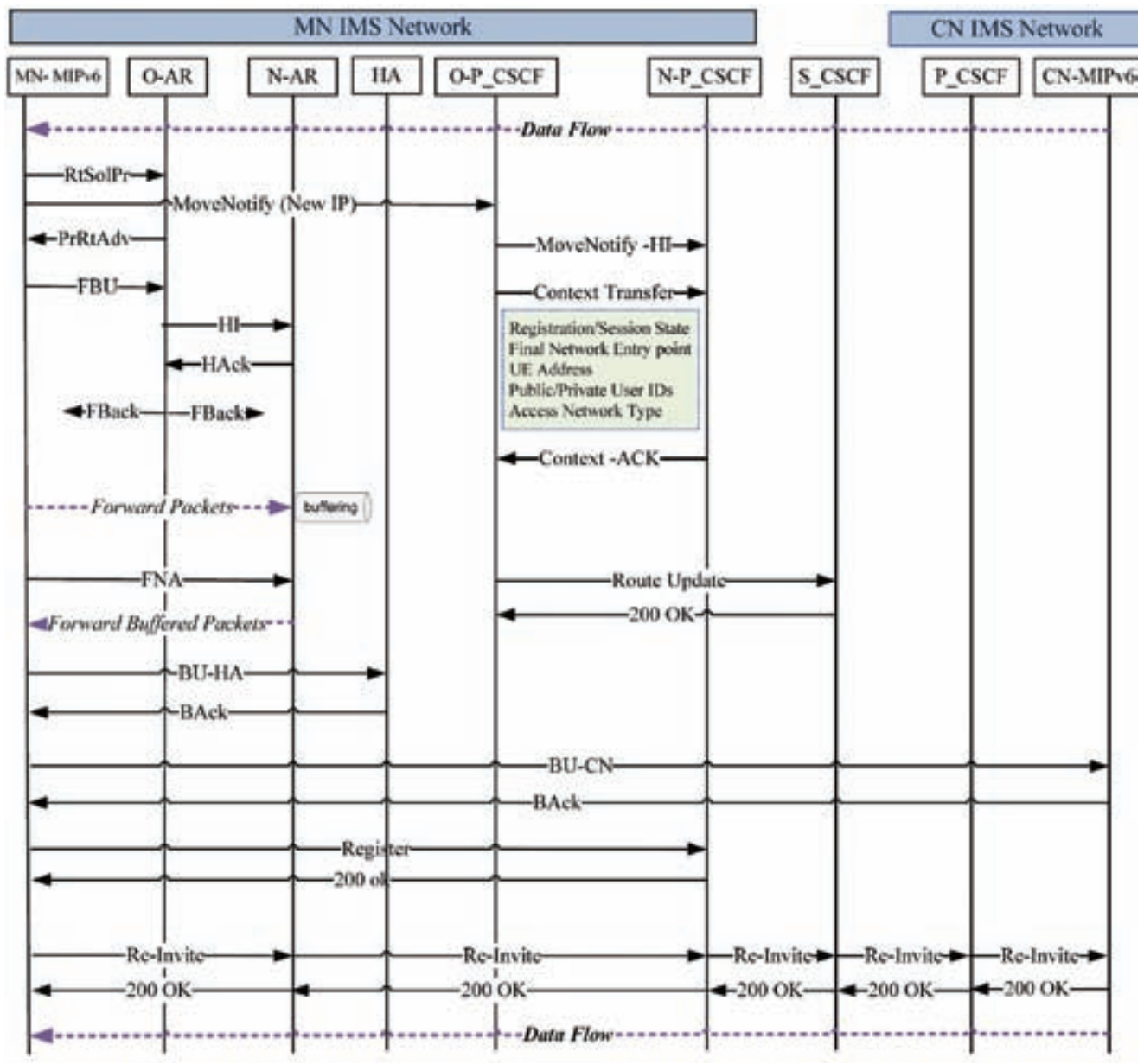

Fig. 5 Proposed Predictive FMIPv6 handover in IMS networks

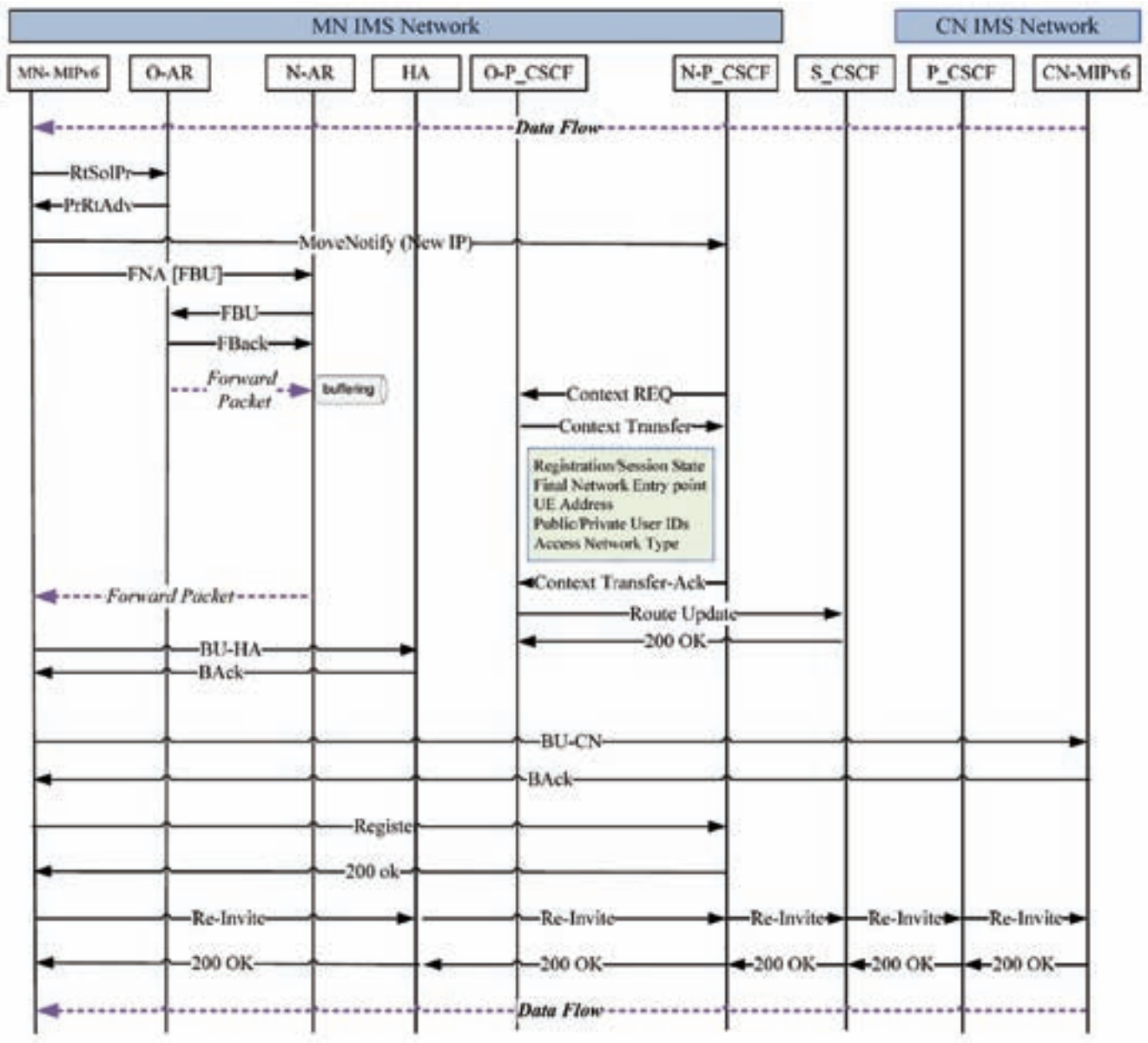

Fig. 7 The proposed Reactive FMIPv6 handover in IMS networks

### 3.4 Proposed scheme II

In this scheme, the MN has performed a handover before the context transfer is requested. It sends a move-notify message to the new P-CSCF containing the old P-CSCF IP address (see Figure 7). The old P-CSCF is the one that validates the transfer. The New P-CSCF sends a Context Transfer Request message to the old P-CSCF. The old P-CSCF replies with a Context Transfer Message that contains the session information. Finally, after receiving Ack, the old P-CSCF updates S-CSCF routes information by sending Route update and receives 200 ok. In this scheme context transfer does not run simultaneously with MIPv6 handover procedure.

The Proposed Reactive FMIPv6 handover Timing Diagram in IMS network is shown in Figure 8.

## IV. PERFORMANCE EVALUATION

In this section we evaluate the handover delay for the two proposed schemes and compare with the delay of the original MIPv6 in IMS. The analysis is based on a simple model presented in [14] that takes into account the delay of the different entities involved in the handover. For simplicity, we consider the model illustrated in Figure 9. Table 1 describes the notations which are used in this model.

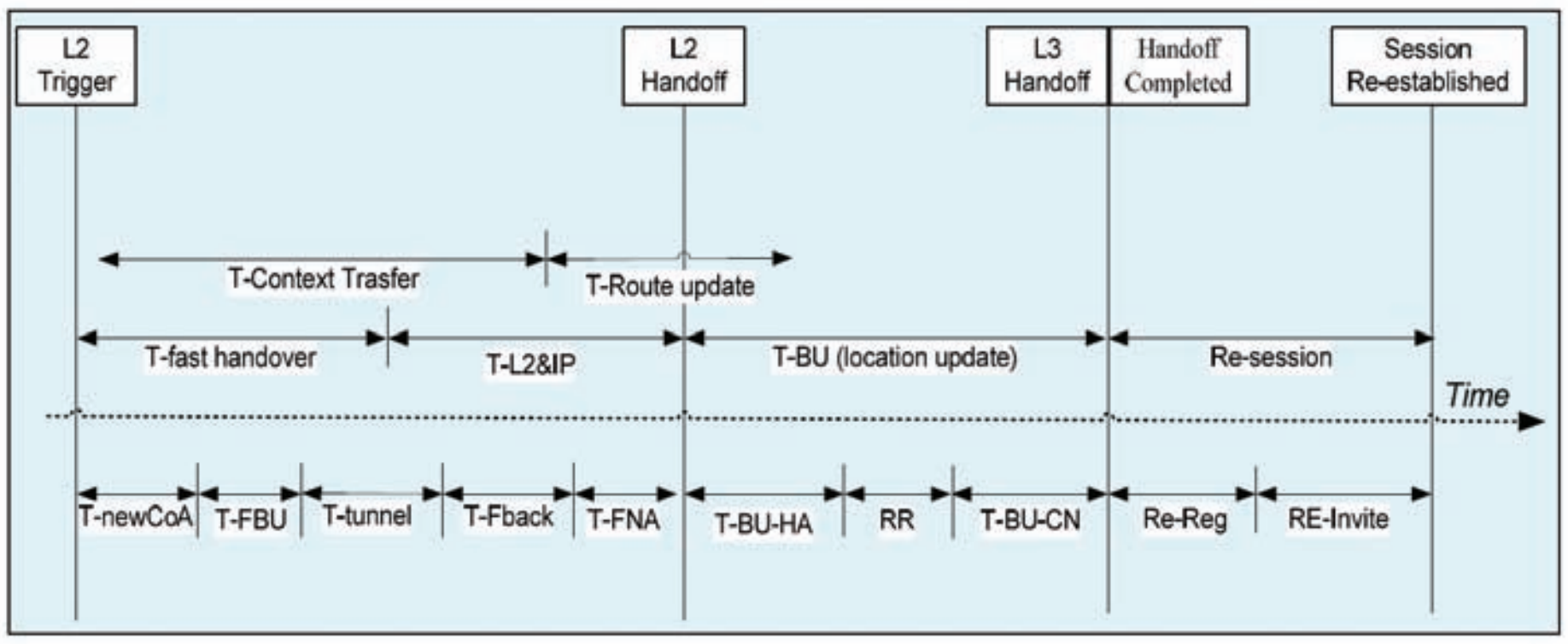

Fig.6 Time diagram for proposed Predictive FMIPv6 handover

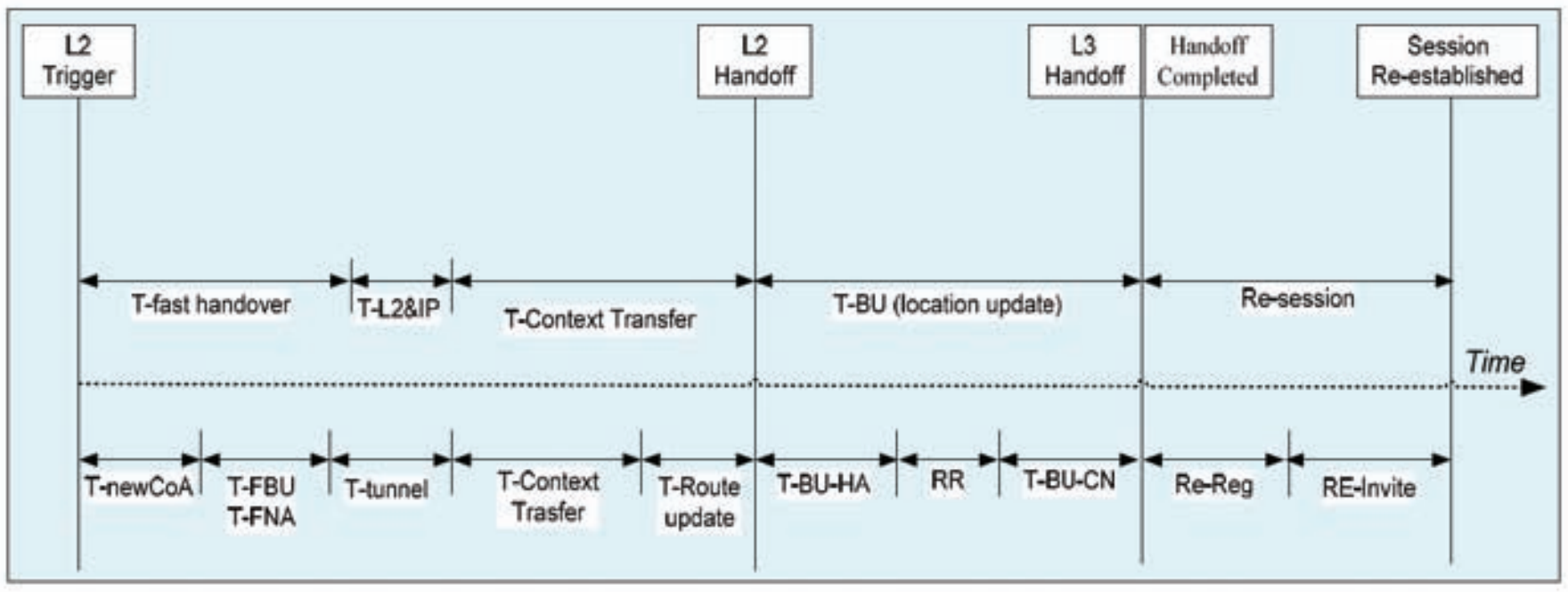

Fig. 8 Time diagram for the proposed Reactive FMIPv6 handover

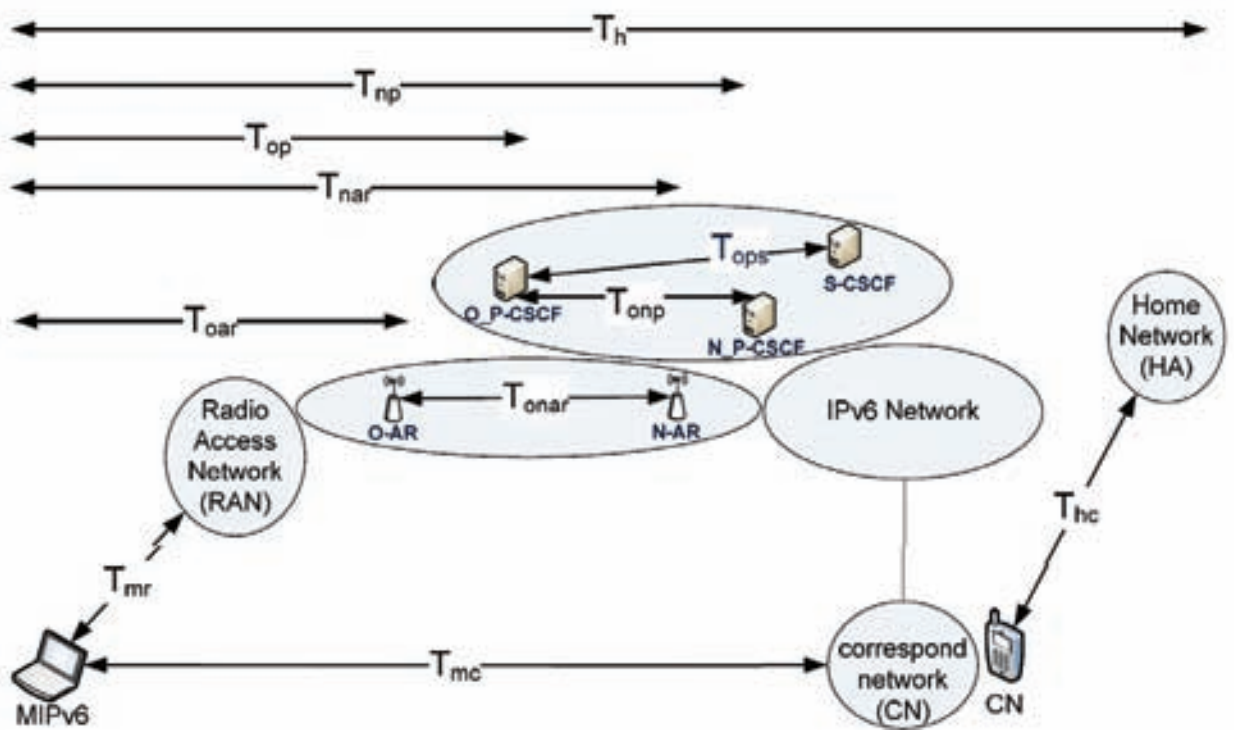

Fig. 9 Simple model for analysis

Table 1 Notations used in Figure 6

| Notations | Description (Delay between) |
|---|---|
| $T_{mr}$ | The MN and the radio access network |
| $T_{oar}$ | The MN and the old AR |
| $T_{nar}$ | MN and the new AR |
| $T_{onar}$ | The OAR and the NAR |
| $T_{op}$ | The MN and the old P-CSCF |
| $T_{np}$ | The MN and the new P-CSCF |
| $T_{onp}$ | The old P-CSCF and the new P-CSCF |
| $T_h$ | The MN and its HA |
| $T_{ops}$ | The old P-CSCF and the S-CSCF |
| $T_{mc}$ | The MN and the CN |

We assume that the delays are symmetric and $T_{mr} < T_h$. We do not consider the time needed by DAD process and Return Routability procedure, also the processing and queuing times are not considered.

In standard IMS-MIPv6, the handover is performed as follows: first the MN sends router solicitation to all ARs in its domain and receives router advertisement from them, then, selects one of them to migrate. Afterward, the MN obtains a new CoA, which takes $2T_{nar}$. Then the MN sends BU to HA and CN and receives Back from them, which takes $2T_h + 2T_{mc}$. Finally the MN should register in the new IMS network, and Re-invite the CN; these procedures take $4T_{np} + 8T_{mc}$. Therefore the overall delay for the standard IMSMIPv6 handover would be:

$$T_{st} = 2T_{nar} + 2T_h + 10T_{mc} + 4T_{np} \qquad (1)$$

In the proposed predictive scheme the handover procedure starts in the same way as for standard scheme with proxy router solicitation/advertisement which takes 2Toar. The MN then sends Move-Notify that contain new MN's IP address, to the old P-CSCF, which takes $T_{op}$. The MN sends an FBU to the old AR, which takes $T_{oar}$. The ARs then exchange HI and HAck, which takes $2T_{onar}$. The previous AR sends an FBAck to the new AR and to the MN, which takes at most $T_{oar}$. At the same time the old P-CSCF, after transmitting Move-Notify, sends context-transfer to the new P-CSCF. After receiving context, the new P-CSCF sends context-transfer ACK to the old P-CSCF. Then the old PC-SCF sends route update to S-CSCF and receives 200 ok. This procedure proceeds concurrently with the handover process avoiding extra latency. Subsequently, the MN sends an FNA to the new AR, which takes $T_{nar}$. Then the MN sends BU to HA and CN and receives Back from them, which takes $2T_{ha} + 2T_{mc}$. Finally the MN should re-register in the new IMS network, and re-invite the CN; these procedures take $2T_{np} + 2T_{mc}$. So the delay for the proposed Predictive IMS-FMIPv6 handover would be:

$$T_{prd} = 4T_{oar} + T_{op} + 2T_{onar} + T_{nar} + 2T_{h} + 4T_{mc} + 2T_{np} \quad (2)$$

In the proposed reactive scheme the handover procedure starts in the same way as standard scheme with proxy router solicitation/advertisement, which takes $2T_{oar}$. The MN then sends Move-Notify that contain new MN's IP address, to the new P-CSCF, which takes $T_{np}$. At this moment the MN has moved to the new AR and has not received an FBAck. It thus sends an FBU encapsulated in an FNA via the new AR to the previous AR. The previous AR sends back an FBAck to the new AR and starts forwarding packets to the new AR, if the new CoA is accepted. This procedure takes $T_{nar} + 2T_{onar}$. Then, the new AR delivers the forwarded packets immediately to the MN. Subsequently, the new P-CSCF sends context request to the old P-CSCF. The old P-CSCF sends session information to the new P-CSCF by context-transfer and receives ACK after. Then the old P-CSCF sends route update to S-CSCF and receives 200 ok. This procedure takes $3T_{onp} + 2T_{ops}$. Afterward, the MN sends BU to HA and CN, and receives Back from them, which takes $2T_{ha} + 2T_{mc}$. Finally the MN should re-register in the new IMS network and reinvite the CN; these procedures take $2T_{np} + 2T_{mc}$. As the result, the delay for the proposed reactive IMS-FMIPv6 handover would be:

$$T_{rac} = 2T_{oar} + 3T_{np} + T_{nar} + 2T_{onar} + 3T_{onp} + 2T_{ops} + 2T_{h} + 4T_{mc} \quad (3)$$

## V. NUMERICAL RESULT

In this section, we present the performance evaluation by simulations over OPNET for the standard and the two proposed schemes. To evaluate the disruption time, we set $T_{mr}$ = 10 ms as in [5] and [13]. Also, we assume $T_{oar}$ = 11ms, $T_{onar}$ = 5ms, $T_{op}$ = 15ms, $T_{onp}$ = 7ms and $T_{ops}$= 10ms. The delay introduced by the Internet depends on the number of routers and the type of links in the path of datagram transmission. For this reason, we suppose the one-way Internet delay over the wired network to be constant, i.e., equals to 100ms. Therefore, the delays are considered as $T_{h}$ = 116ms, $T_{mc}$ = 128ms, and $T_{hc}$ = 114ms.

In Figure 10, it can be noticed that the disruption time for the standard scheme gets larger when the delay increases. The proposed predictive scheme gives slightly shorter handover delay comparing with the proposed reactive scheme.

In Figure 11, the handover delays for the proposed schemes are considerably lower than the standard scheme when the delay between the MN and the home network increases.The Figure 12 shows that as the delay between the old P-CSCF and the new P-CSCF isincreasing, the disruption time for the proposed predictive scheme would be the lowest.This augmentation is enhanced for the higher values of the old to new P-CSCF delay.

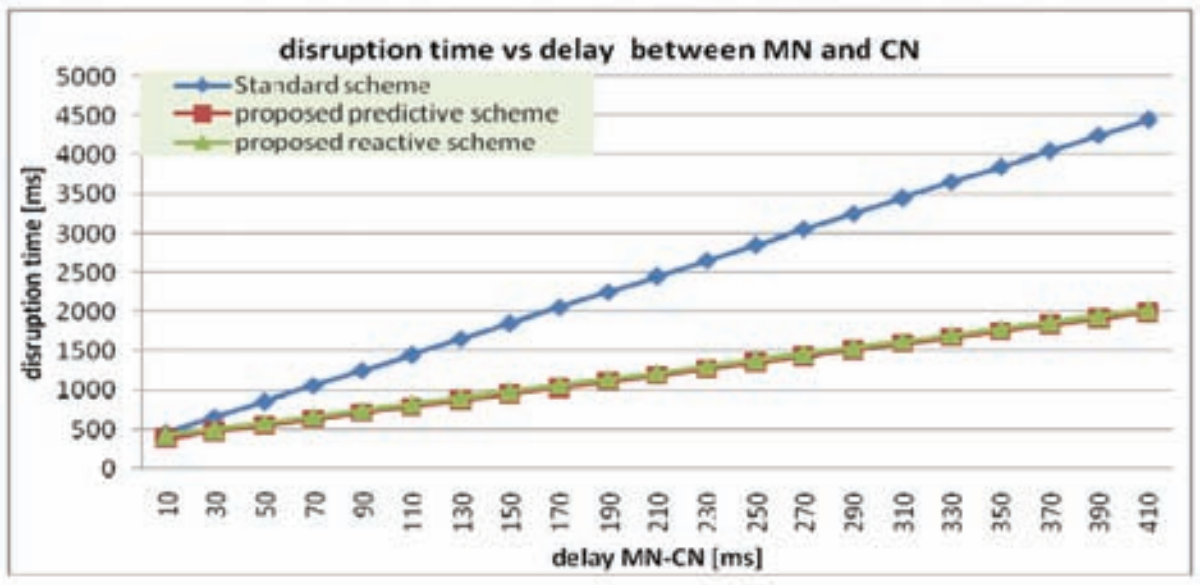

Fig. 10 Disruption time versus delay between MN and CN

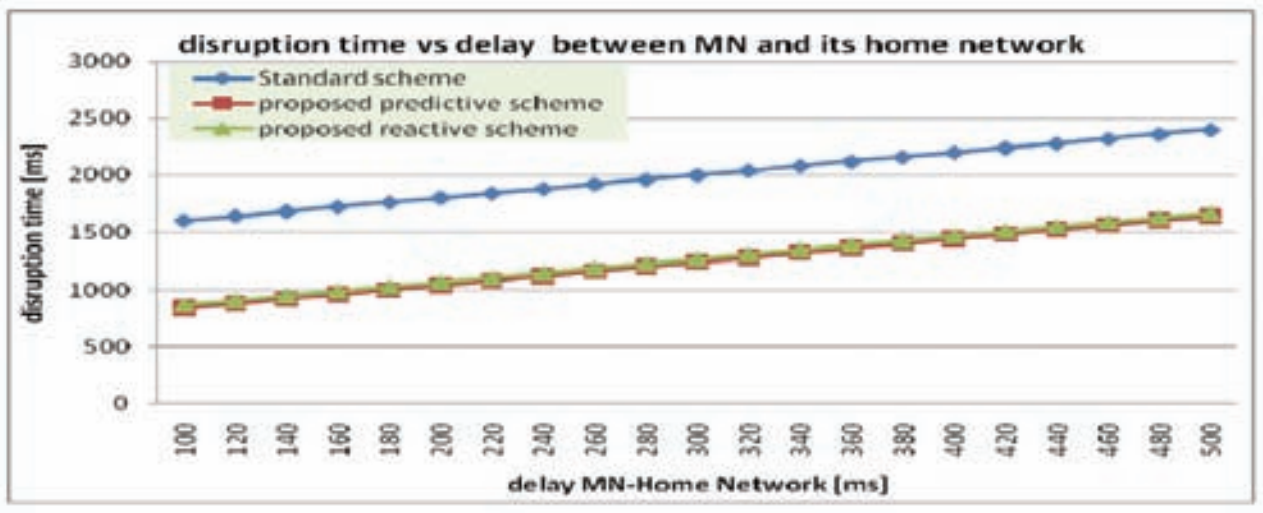

Fig.11 Disruption time versus delay between MN and HA

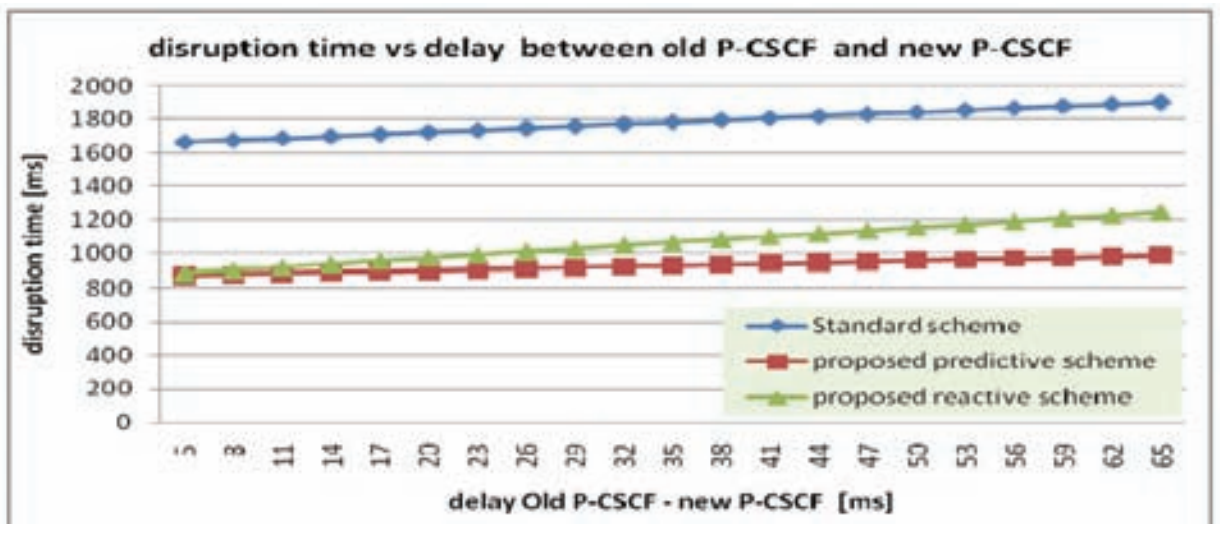

Fig.12 Disruption time versus delay between the old and the new P-CSCFs

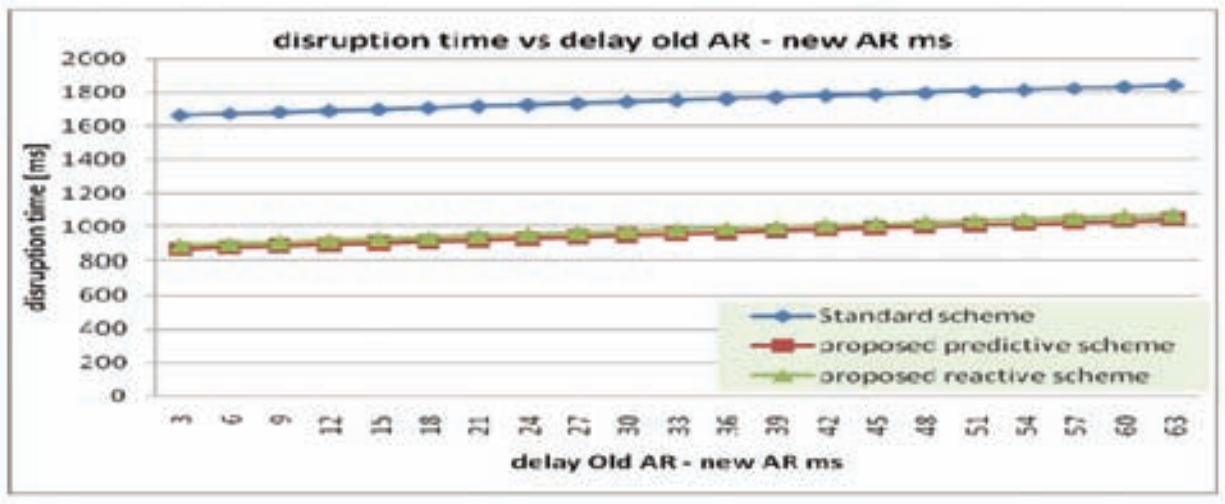

Fig 13 Disruption time versus delay between the old and the new ARs

Moreover, this figure adopts that the Disruption time rises monotonically as the old to the new P-CSCF delay increases.

Figure 13 illustrates that the relation between the disruption times and the old to new AR delay. Similarly, by increasing the delay between the old AR and the new AR, disruption time increases for all the three schemes, however, the proposed schemes have better performance comparing to the standard one.

The proposed schemes make it possible to have a handover between P-CSCFs, without losing SIP state information. Additionally, fewer messages (especially in predictive scheme) are used for the Re-authorization of the MN, and for the re-establishment of the session, which implies shorter handover delay as compared to the standard scheme.

## VI. QOS PROVISIONING IMPROVEMENT

In order to investigate handover with QoS negotiation, we consider the IMS-SIP signaling components and methods for QoS provisioning. We previously showed how the latency during handover between P-CSCFs can be reduced by context transfer. In this section, we show that instead of re-negotiating the session parameters needed for QoS provisioning, the previous session information, saved by CSCF server, can be used.

### 6.1 QoS provisioning in IMS

As mentioned before, The IMS allows users to benefit from resource consuming multimedia services over unified transport. It provides end-to-end QoS using capabilities of access and transport networks. In IMS, an UE negotiates its parameters and QoS requirements during a SIP session establishment or modification procedure. After negotiating the QoS parameters at the application level, the UE reserves appropriate resources from the access network. To guarantee the required QoS in the interconnecting backbone, IMS assumes that operators negotiate service-level agreements. The IMS QoS mechanism is largely based on the interaction between IMS and the underlying access network. When an UE hands off among IMSs over heterogeneous access networks, the QoS parameters have to be re-negotiated between the newly visited IMS and its underlying access network. The delay of resource reservation along the newly established data path during the UE's movement may cause service disruption for real-time services.

In the IMS standard[15], it is specified that QoS parameters can be negotiated between two UE's prior to the session establishment. Once the QoS parameters have been negotiated (checked against network resources and constraints) and been approved/modified by the CSCF (checked against user subscription credentials), the IMS network asks the CN and the access network to reserve resources for this session. The SIP Invite and SIP ACK messages are used for this purpose as shown in Figure 14.

The first Invite message from the MN to the CN carries the QoS proposal (request) of the sender and this proposal is checked against the subscription levels of users at the P-CSCF and S-CSCF in the home networks of both MN and CN, and QoS parameters are modified at these locations if there is a mismatch. Afterwards, UE2 puts her own QoS proposal in the answer and this proposal is again checked and modified at the associated S-CSCF's and P-CSCF's of both users according to their subscription status. Finally, MN can accept this counter QoS proposal and start the session or try to renegotiate with a SIP ACK message.

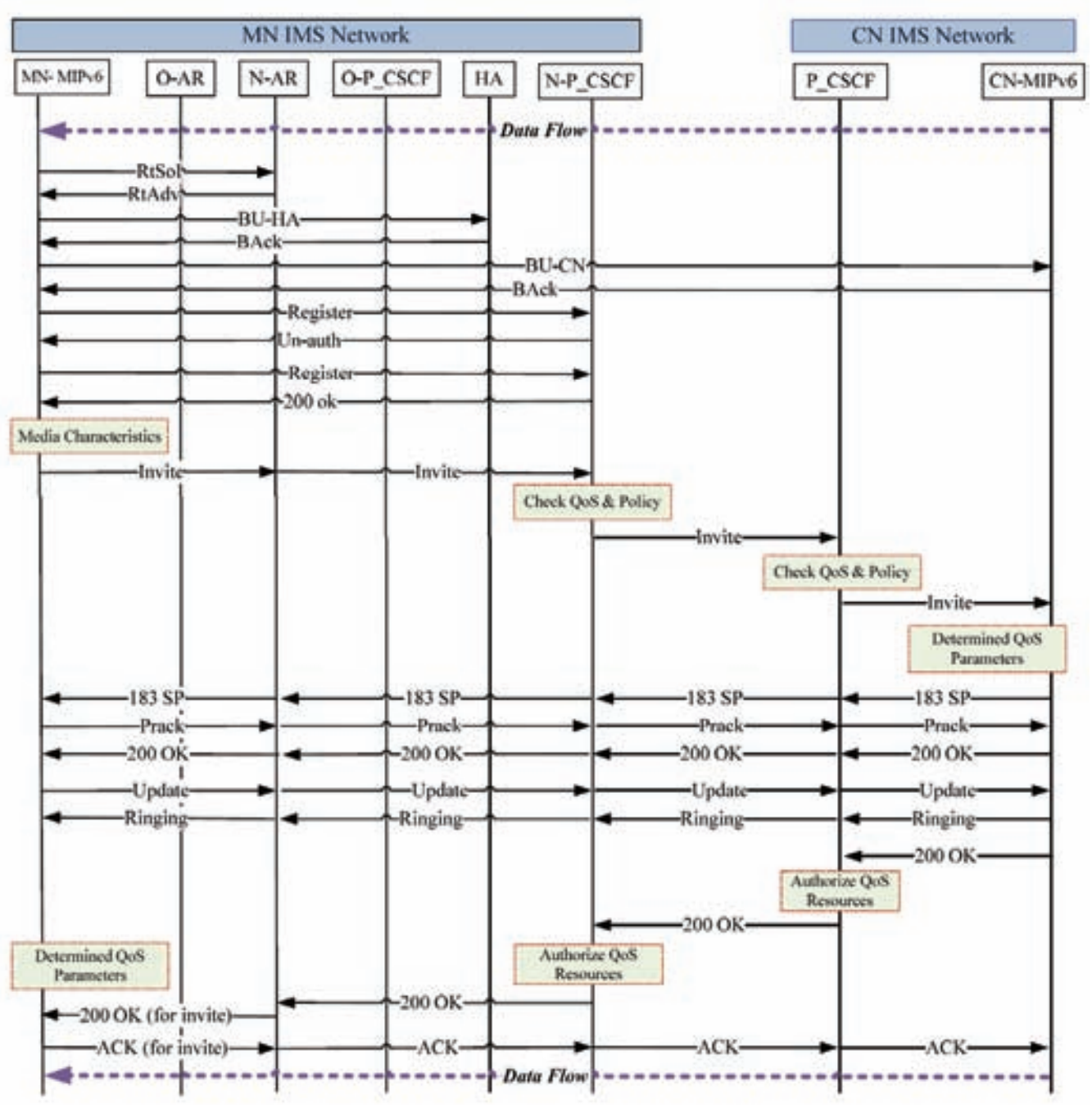

Fig.14 Standard QoS in IMS

To follow the resource availability while maintain the service quality, QoS parameters are continuously negotiated and compromised between IMS networks. Over provisioning can be employed for less resource demanding services such as the legacy SMS messaging, with only 160 bytes in size and no real-time requirements. However, for multimedia components, such as video streaming, high data throughput and low signaling delay are critical at the network side, while high processing power, memory-storage space and energy are required at the client side until the end of the session.

**6.2 Proposed predictive context tranfer scheme for QoS provisioning**

In Figure 15 we have shown our Predictive Context Transfer-based scheme for QoS Provisioning. As mentioned before, in this scheme the MN is aware that toward which access router it will be heading and anticipate the new P-CSCF in advanced. After sending a movement notification by the new P-CSCF to the old P-CSCF, the old P-CSCF sends a movement notification to the new one. When the old AR catches the MN's movement, it sends QoS context, acquired from the old P-CSCF, to the new P-CSCF and S-CSCF. After the session context transfer is completed, the new P-CSCF sends QoS context to the new AR. Following this procedures, the MN sends Re-Invite message to the CN's P-CSCF and the new P-CSCF.

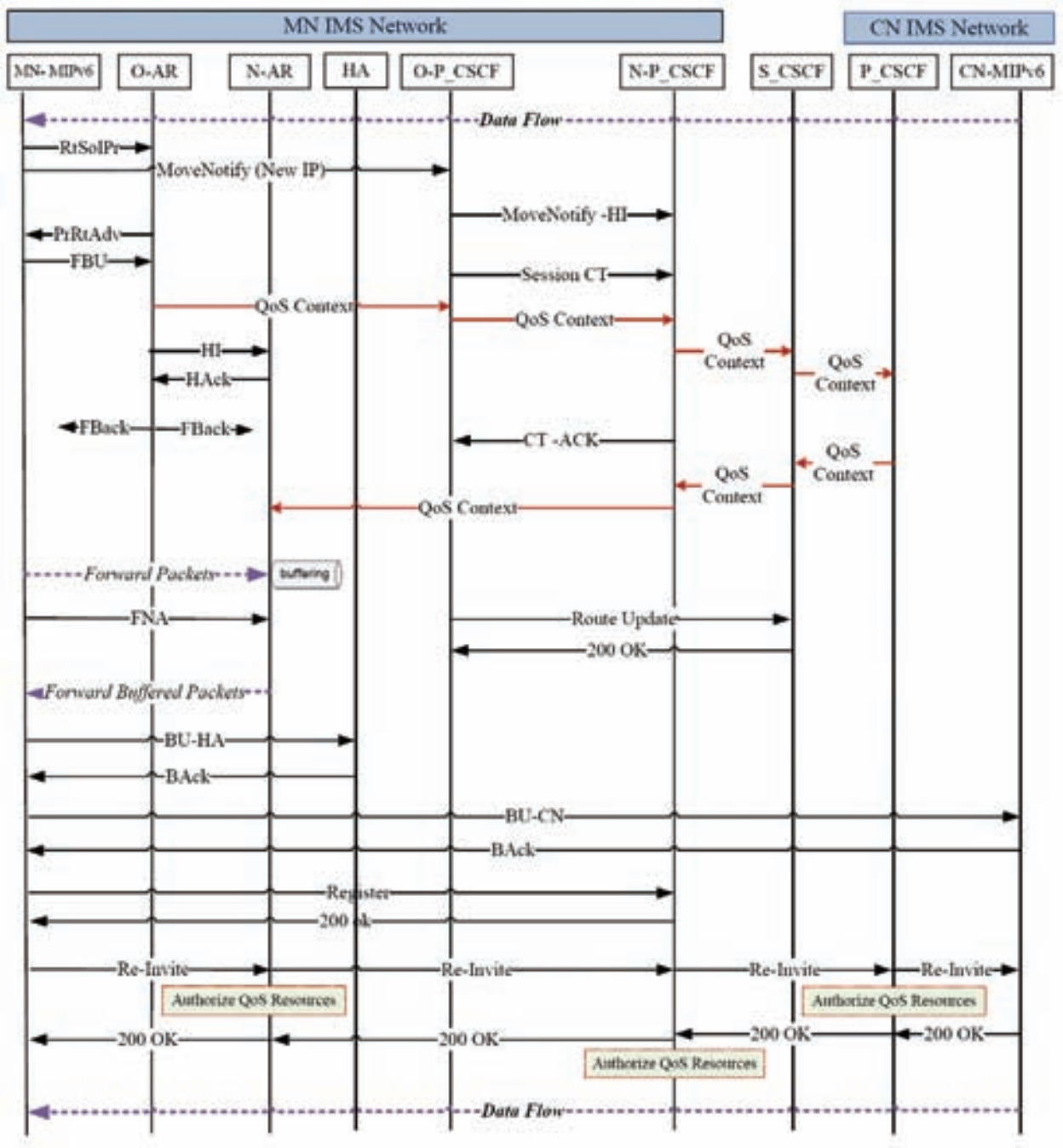

Fig. 15 Proposed QoS Context Transfer scheme based on Predictive FMIPv6 handover Scheme

**6.3 Proposed reactive context transfer scheme for QoS provisioning**

Our proposed Reactive Context Transfer-based scheme is illustrated in n Figure 16. In this scheme the MN has performed a handover before the context transfer is requested. After sending a movement notification to the new P-CSCF and creating a tunnel between the old and the new ARs, the new P-CSCF sends request massages for session and QoS context to the old P-CSCF. The old P-CSCF receives QoS context from the old AR after sending the request to it. Upon receiving QoS context, the old P-CSCF forwards it to the new S-CSCF and NAR. Subsequently, the MN sends Re-Invite message to the CN's PCSCF and the new P-CSCF.

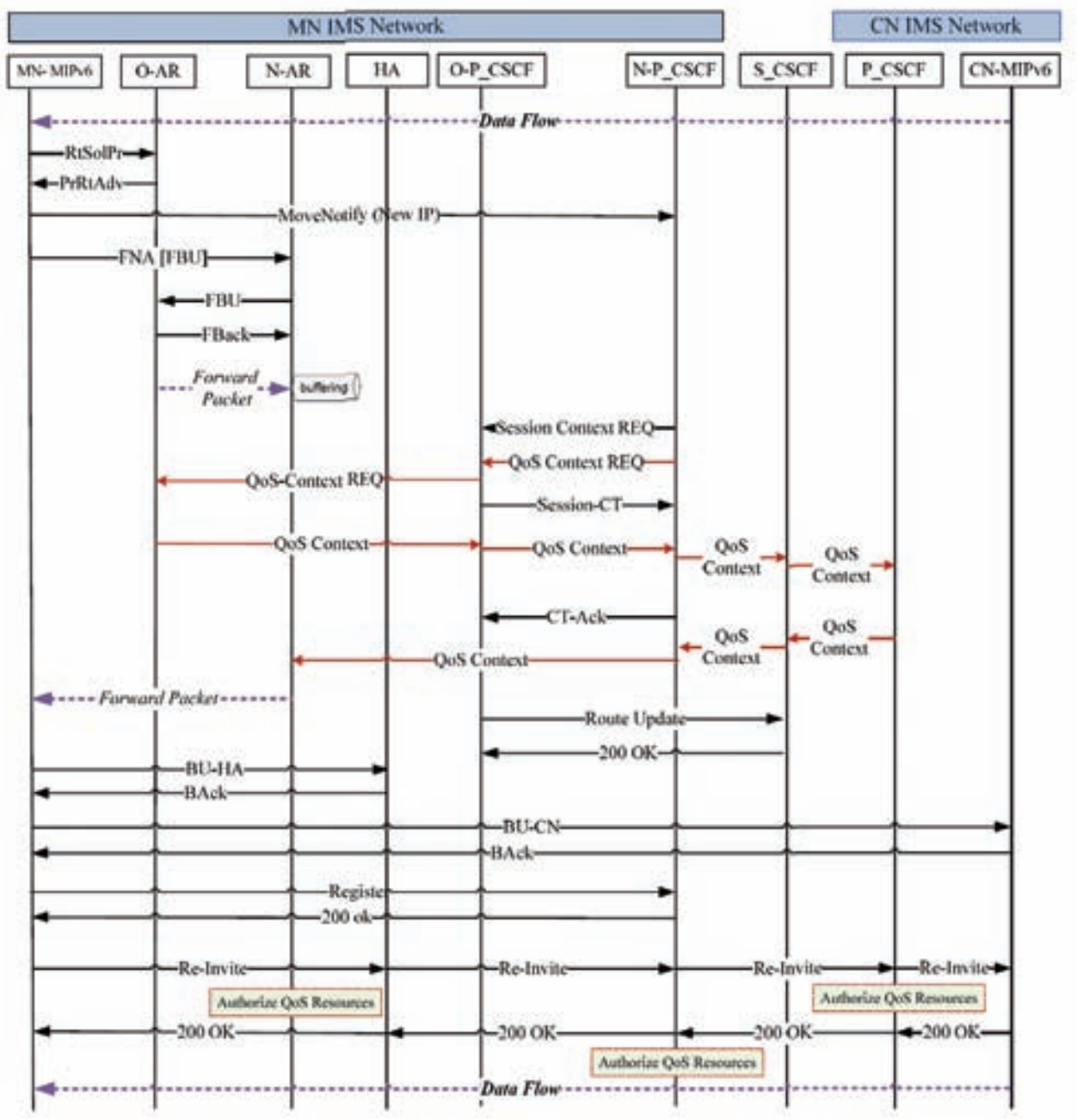

Fig 16 Proposed QoS context transfer scheme based on Reactive FMIPv6 handover scheme

### 6.4 Performance evaluation

In this subsection we compare the handover delay for our two proposed schemes for handover with QoS negotiations. As mentioned before, in the Predictive scheme, the session and QoS context transfer procedures are performed simultaneously to MIPv6 handover and don't cause excessive latency. However, in reactive scheme the period needed for QoS negotiations should be added to the total latency.

The simulation performance evaluation for the standard and the two proposed schemes, illustrated in Figure 17, shows that the predictive scheme cause less service disruption duration comparing to the reactive scheme, while performance of the reactive scheme is much better than the standard method.

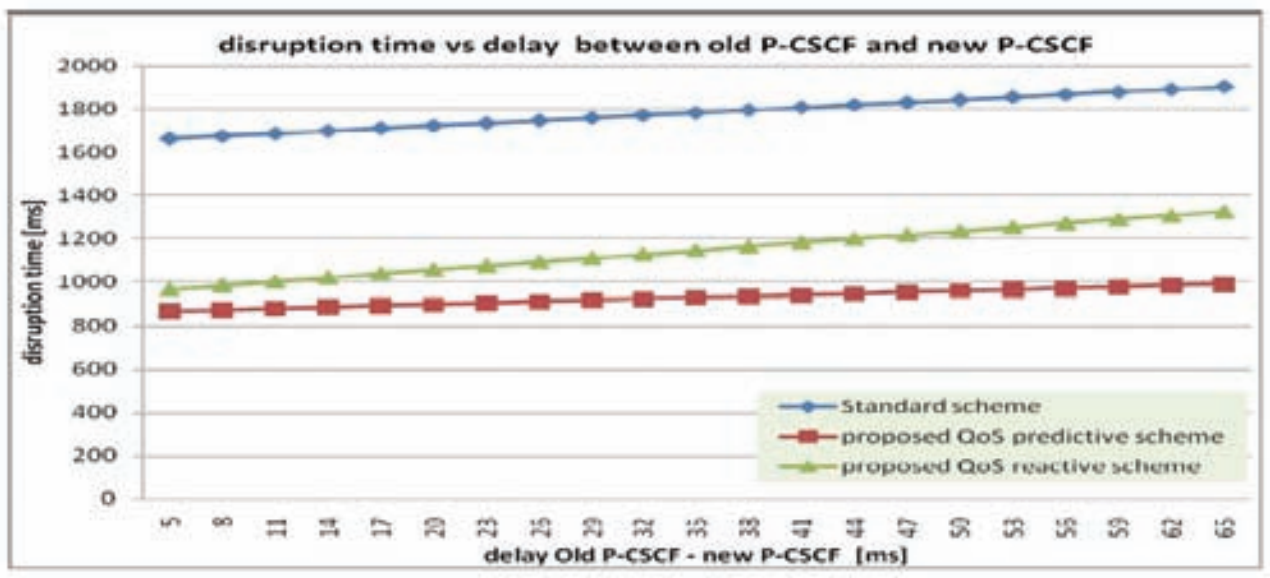

Fig.17 Disruption time versus delay between the old and the new P-CSCFs

## VII. CONCLUSION

In this paper, In order to improve the service quality of IMS over MIPv6, we proposed two methods for handover between P-CSCFs. To lessen the handover latency, the predictive and reactive proposals make use of context transfer between the old and the new P-CSCFs. Transferring the session state information prior to the handover, decreases the total time needed for the handover procedure. We investigated the performance of these schemes and compared them with the standard IMS handover without context transfer. Our performance evaluation implies that shorter handover latency is achieved as fewer messages are required for the re-register and re-invite processes, and the session reestablishments. Furthermore, to reduce the service disruption during UE's movement along the new data path, we presented two QoS context transfer schemes based on our two proposed handover mechanisms. Our numerical comparison showed that the context transfer for QoS provisioning causes less service disruption for predictive scheme over reactive one, albeit these two proposed schemes have much better performance comparing to the standard method.

## References

[1] 3GPP TS 23.228 V8.3.0. IP Multimedia Subsystem (IMS); Stage 2. Jan. 2007.
[2] Johnson D. B., Perkins C. E., and Arkko J. Mobility support in IPv6. IEFT RFC 3775, June 2004.
[3] Koodli R. Mobile IPv6 Fast Handovers. IETF RFC 5268, June 2008.
[4] Soliman H., Castelluccia C., El-Malki K., and Bellier L. Hierarchical Mobile IPv6 (HMIPv6) Mobility Management. IETF RFC 5380, Oct. 2008.
[5] Kempf J. Problem Description: Reasons for Performing Context Transfers Between Nodes in an IP Access Network. IETF RFC 3374, September 2002.
[6] Loughney J., Nakhjiri M., Perkins C. and Koodli R. Context Transfer Protocol (CXTP). IETF RFC 4067, July 2005.
[7] Lynggaard K., Vestergaard E., Schwefel H-P. and Kuhn G. Optimized Macro Mobility within the 3GPP IP Multimedia Subsystem. IEEE, ICWMC 2006.
[8] Renier T., K. Larsen L., Castro G. and Schwefel H-P. Mid-Session Macro-Mobility in IMS-based networks. IEEE, Vehicular Technology Magazine March 2007.
[9] Koodli R., Perkins C. E. Fast Handovers and Context Transfers in Mobile Networks. ACM SIGCOMM Computer Communication Review October 2001; 31(5).
[10] Liu C., Qian D., Liu Y., and Xiao K. A Framework for End-to-End QoS Context Transfer in Mobile IPv6. PWC 2004, pp. 466–475, LNCS 3260.
[11] 3GPP TS 24.228, V5.15.0. Signaling flows for the IP multimedia call control based on SIP and SDP; Stage 3 (Release 5). June 2006. 18
[12] Narten T., Nordmark E. and Simpson W. Neighbor Discovery for IP Version 6. IETF RFC 2461, Dec. 1998.
[13] Costa X., Torrent-Moreno M., and Hartenstein H. A Performance comparison of mobile IPv6, hierarchical mobile IPv6, fast handovers for mobile IPv6 and their combination. IEEE, Mobile Comp. Commu. Rev., Oct. 2003; 7(4): 5–19.
[14] Fathi H., Chakraborty S., and Prasad R. Optimization of Mobile IPv6-Based Handovers to Support VoIP Services in Wireless Heterogeneous Networks. IEEE Transactions on Vehicular Technology Jan. 2007; 56(1).
[15] 3GPP TS 29.208 V6.7.0. End-to-end Quality of Service (QoS) signalling flows (Release 6). Jun. 2007.

## Biography

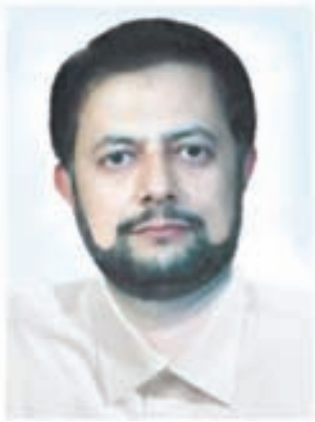

*Reza Farahbakhsh* received his B.Sc. in 2006. He is currently an M.S. student at the department of computer engineering, Isfahan University, Isfahan, Iran.

*Naser Movahhedinia* received his B.Sc. from Tehran University, Tehran, Iran in 1967, and his M.S. from Isfahan University of Technology, Isfahan, Iran in 1990 in Electrical and Communication Engineering. He got his Ph.D. degree from Ottawa- Carleton Graduate Institute, Ottawa, Canada in 1997, where he was a research associate at System and Computer Engineering Department, Carleton University for a short period after graduation. Currently he is an assistant professor at Computer Department, The University of Isfahan. His research interests are wireless networks, personal communications systems and Internet Technology.

(Continued from page 143)

## Biography

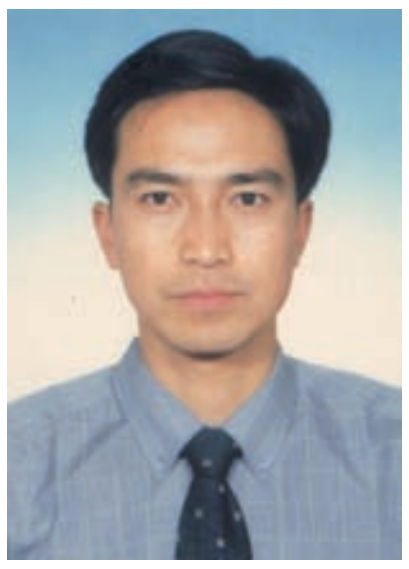

*Ping Yi* was born in 1969. Currently he is Associate Professor at School of Information Security Engineering, Shanghai Jiao Tong University in China. He received the BSc degree in department of computer science and engineering from the PLA University of Science and Technology, Nanjing, in 1991. He received the MSc degree in computer science from the Tongji University, Shanghai, in 2003. He received the Ph.D degree at the department of Computing and Information Technology, Fudan University, China. His research interests include mobile computing and ad hoc networks security. He is Project Leader for Intrusion Detection and active protection in Wireless ad hoc Networks, supported by the National High-tech Research and Development Plan of China from 2007 to 2009. He is a member of IEEE Communications and Information Security Technical Committee, Associate Editor for Wiley's Security and Communication Networks (SCN) Journal, Editor for Journal of Security and Telecommunications, Technical Program Committee (TPC) for the ICC'09 CISS (ICC 2009 Communication and Information Systems Security Symposium), the GC'08 CCNS (IEEE Globecom 2008 Computer and Communications Network Security Symposium) .